\documentclass[pre,amssymb,superscriptaddress,floatfix,a4paper,twocolumn,showpacs]{revtex4-1}
\usepackage{graphicx}
\usepackage{bm}
\usepackage{amsmath}
\usepackage{amsthm}
\usepackage[normalem]{ulem}
\usepackage[usenames]{color}

\begin{document}

\title{Dynamical heterogeneities as fingerprints of a backbone structure in Potts models}

\author{E. E. Ferrero}
\affiliation{CONICET, Centro At{\'{o}}mico Bariloche, 8400 San
Carlos de Bariloche, R\'{\i}o Negro, Argentina}

\author{F. Rom\'a}
\affiliation{Departamento de F\'{\i}sica, Universidad Nacional
de San Luis, Instituo de F\'{\i}sica Aplicada (INFAP),\\ CONICET,
Chacabuco 917, D5700BWS, San Luis, Argentina}

\author{S. Bustingorry}
\affiliation{CONICET, Centro At{\'{o}}mico Bariloche, 8400 San
Carlos de Bariloche, R\'{\i}o Negro, Argentina}

\author{P. M. Gleiser}
\affiliation{CONICET, Centro At{\'{o}}mico Bariloche, 8400 San
Carlos de Bariloche, R\'{\i}o Negro, Argentina}

\date{\today}

\begin{abstract}

We investigate slow non-equilibrium dynamical processes in two-dimensional $q$--state Potts model with both ferromagnetic and $\pm J$ couplings. Dynamical properties are characterized by means of the mean-flipping time distribution. This quantity is known for clearly unveiling dynamical heterogeneities. Using a two-times protocol we characterize the different time scales observed and relate them to growth processes occurring in the system. In particular we target the possible relation between the different time scales and the spatial heterogeneities originated in the ground state topology, which are associated to the presence of a backbone structure. We perform numerical simulations using an approach based on graphics processing units (GPUs) which permits to reach large system sizes.  We present evidence supporting both the idea of a growing process in the preasymptotic regime of the glassy phases and the existence of a backbone structure behind this processes.

\end{abstract}


\maketitle

\section{Introduction}
\label{sec:intro}

It has became clear during the past years that when studying the general class of systems with slow dynamics, the heterogeneous character of both spatial and dynamical properties plays a fundamental role~\cite{lesuch1,berthier_book,berthier_biroli_rmp}. For instance, the increasing interest in the heterogeneous behavior of glasses and related complex systems during the past decade is concomitant with the fact that its understanding is of great significance for a complete comprehension of the glass transition problem. In this direction, both dynamical and spatial heterogeneities have been studied in several systems such as colloids~\cite{Kegel2000,Weeks2000,Cui2001,Ho2004,Kaufman2006,Hunter2012}, granular matter~\cite{Dauchot2005,Ferguson2007,Reis2007,Pica2007,Watanabe2008,Fiege2009,Candelier2009,Berardi2010}, structural~\cite{Bennemann1999,Berthier2005,appignanesi2006,Castillo,Mazoyer2009} and spin glasses~\cite{Chamon02,Castillo02,Castillo03,Chamon04,Montanari03a,Montanari03b}.

An important question concerns the key relation between dynamical and spatial heterogeneities. 
Studies in structural glasses revealed that a relation clearly exists,  and much effort has been devoted to identify the involved mechanisms that characterize it. In contrast, spin glasses have shown to be more elusive regarding possible links between spatial and dynamical heterogeneities. It has recently been shown that a direct connection between dynamical and spatial heterogeneities is present in the $\pm J$ Edwards-Anderson (EA) model~\cite{EA}, both in two and three dimensions~\cite{Ricci,Roma2,Roma3,roma2010}.

Dynamical heterogeneities in the low temperature dynamics of spin-glass models can be characterized through the mean flipping-time distribution (MFTD)~\cite{Ricci}, which gives information on the amount of spins flipping events within a given time window. 
For the two-dimensional $\pm J$ EA model, where a spin-glass phase does not exist ($T_g=0$) but preasymptotic slow dynamics can be observed at low temperatures, it has been shown that this distribution develops two characteristic peaks when the temperature is decreased~\cite{Roma2}. 
There is a first temperature-independent peak at short time scales characterizing fast degrees of freedom, and a second peak at large time scales which is thermally activated and is related to the slow-degrees of freedom.
This strong dynamical heterogeneity has been related to the underlying spatial heterogeneities given by the backbone structure of the model~\cite{Roma2,Roma3}. 
The backbone structure is a constrained structure fully characterized by the topological properties of the ground state. 
In particular, for the $\pm J$EA model, an analysis of the degenerate configurations of the ground state reveals different sets of spins: on the one hand solidary spins which form a ferromagnetic-like state and on the other hand non-solidary spins which can be seen as paramagnetic-like. 
This property naturally links to the information obtained through the MFTD: non-solidary (paramagnetic-like) and solidary (ferromagnetic-like) spins can be recast as responsible of the fast and slow degrees of freedom, respectively.

The fact that solidary spins have a ferromagnetic-like character has been reinforced by information obtained with the three-dimensional $\pm J$ EA model. 
In this case there exists a spin-glass phase at low temperatures (below $T_g=1.12$~\cite{katzgraber2006}) and again the strong dynamical heterogeneities observed in the MFTD can be directly related to the backbone structure obtained using ground state information~\cite{roma2010}. 
More importantly, it has been shown that following only the set of solidary spins a domain growth process can be identified~\cite{roma2010}. 
The observed domain growth dynamics intuitively suggests the idea of a characteristic growing length in the system. 

The key obstacle in these studies concerns the identification of the backbone structure, which is limited by the difficulty to properly identify ground state configurations. 
Given the high degeneracy of the ground state of the $\pm J$ EA model, this sets a finite size constraint, and thus only systems with very small sizes can be resolved. 
At the same time, the definition of the backbone structure for the $\pm J$ EA model is based on the degeneracy of the ground state and on the Ising character of the spin variable. A generalization to systems with only a simple degenerated ground state (i.e. with up-down symmetry), such as the Gaussian model, is far from being straightforward~\cite{roma2010b,roma2012p}. The generalization to non-Ising spin systems, such as XY or Potts models, is also an important open issue. 

In this work, in order to gain insight into the ingredients which should be taken into account for a generalization of the definition of the backbone structure, we first analyze the relation between dynamical and spatial heterogeneities in the two-dimensional Potts model with ferromagnetic couplings.
This model appears as a good candidate: it has a ground state with a very low degeneracy with only $q$ ground state configurations and it also has a very rich low temperature dynamics with many qualitatively different regimes~\cite{ferrero2007,loureiro2012}. 
Furthermore, new numerical computation platforms allows for large system sizes to be reached, reducing finite size effects and also allowing for very long time regimes~\cite{ferrero2012}. 
Moreover, we extend our analysis and also consider a frustrated version of the Potts model, the $\pm J$ Potts model in two dimensions, which can be regarded as a generalization of the $\pm J$ EA model to $q$ available states for each spin. 
Our analysis reveals evidence of the presence of a backbone structure in this model. 

The outline of the paper is as follows. In Sec.~\ref{sec:model} we define the models we are interested in, i.e. the Potts model with ferromagnetic and $\pm J$ couplings. We also define the MFTD used in this work to characterize dynamical heterogeneities and give detailed information on numerical implementations based on graphics processing units. 
Then, in Sec.~\ref{sec:coarsening}, studying the Potts model with $q=9$ and ferromagnetic couplings we show that the MFTD is directly related to the underlying coarsening process.
Section~\ref{sec:pmj} is devoted to the two-dimensional $\pm J$ EA model, where we show that the MFTD suggests the presence of a domain growth process even in the preasymptotic low temperature regime ($T>T_g=0$). 
Finally, in Sec.~\ref{sec:potts-glass} we analyze the $\pm J$ Potts model for different values of $q$ and present results that reveal the presence of a backbone structure in this case. 
Section~\ref{sec:conc} is devoted to a discussion of the presented results.

\section{Models and GPU-based numerical implementation}
\label{sec:model}

\subsection*{The Potts model}
The Hamiltonian for the $q$--state Potts model is~\cite{wu_potts_review}
\begin{equation}
 H = -J \sum_{(ij)} \delta(s_i,s_j),
\end{equation}
where $J>0$, $s_i=1,2,...,q$, the sum runs over pairs of nearest-neighbors sites on the square lattice of linear dimension $L$ and $N=L^2$ sites, and $\delta(s_i,s_j)$ is the Kronecker delta function~\footnote{notice that in order to directly compare with the standard Ising Hamiltonian $H=-J^{\text{Ising}} \sum_{nn} \sigma_i\sigma_j$, with $\sigma_i=\pm 1$ states, energy and temperature scales should be converted using $J=2J^{\text{Ising}}$. The same applies for the relation between the $\pm J$ Potts model and the Edwards-Anderson model.}.
This Hamiltonian favors ferromagnetic states with two neighbors minimizing its energy when their spin variables $s_i$ are in the same state $q$; the energy of the pair being $-J$. 
The energy of a frustrated bond, i.e. $s_i \ne s_j$, is always zero and therefore the energy of the first excitation over the ground state is $4J$. 
The transition between the paramagnetic high-temperature state and the ferromagnetic low-temperature state, in two dimensions, is a second order transition for $q \le 4$ and a first order transition for $q > 4$~\cite{wu_potts_review}. 
The transition temperature in the square lattice is exactly known, and is given by $T_c = 1/ \ln(1+\sqrt{q})$.

Recently, it has been shown that the dynamical behavior at low temperatures presents a rich variety of regimes~\cite{ferrero2007}, which can be ordered in a temperature scale. 
These regimes can be observed after a sudden quench from the high temperature paramagnetic state (typically a fully disordered state corresponding to $T = \infty$) to different working temperatures $T<T_c$. 
If the working temperature is below but close to $T_c$ the dynamics is governed by nucleation events, just up to $T_n < T_c$. 
Below $T_n$ and above a temperature $T_{co}$ coarsening phenomena rules the dynamics, signaled by a power-law decaying energy relaxation. 
Below $T_{co}$, blocked states (stripes or honeycomb-like configurations) preempts fully development of coarsening dynamics.
This regime is therefore characterized by coarsening at intermediate stages of the dynamics and blocked states at late stages, which when occurring dominate the relaxation process. 
For $q>4$, at smaller temperatures (below some temperature $T_g$) the coarsening relaxation is interrupted by a so called \emph{glassy} state~\cite{deOliveira2004,deOliveira2004a,ferrero2007,deOliveira2009}. 
This glassy regime is characterized by dynamical frustration in such a way that the infinite-time and zero-temperature limit state has an excess energy with respect to the ground state. 
At finite temperature when exiting from the glassy state the system can get stuck again in a blocked state. So, below $T_g$ glassy and blocked states govern the dynamics. 
The relative range of the observed dynamical regimes depends on $q$. 
For example, for $q=9$ one has that the critical temperature is $T_c=0.72134752 J$, the nucleation temperature is $T_n \approx 0.718 J \approx 0.995 T_c$, the coarsening temperature is $T_{co} \approx 0.6 J \approx 0.832 T_c$, and the glass temperature is $T_g \approx 0.2 J\approx 0.277 T_c$~\cite{ferrero2007}.

\subsection*{The $\pm J$ Potts model}
In the case of the $\pm J$ Potts model we use the following Hamiltonian
\begin{equation}
\label{eq:H-Potts-glass}
 H = - \sum_{(ij)} \left[ \delta(J_{ij},1)\delta(s_i,s_j) + \delta(J_{ij},-1)(1-\delta(s_i,s_j)) \right],
\end{equation}
where the $J_{ij}$ are $1$ or $-1$ with equal probability; $J_{ij}=1$ favors a  ferromagnetic state (as in the non-disordered case), while $J=-1$ favors unequal neighbor states ($s_i \ne s_j$). 
In this model the ground state is multiply-degenerated with the number of configurations exponentially growing with the system size. 
Mean field studies suggest that this model develops ferromagnetic order at finite temperatures~\cite{de_santis1995}. However, as discussed in Ref.~\cite{janus2009}, there is no evidence for a ferromagnetic phase in finite dimensional models.
Although in three dimensions this system has a low temperature spin-glass phase, with a glass transition temperature decreasing with increasing $q$~\cite{janus2009,janus2010}, the two-dimensional counterpart has $T_g=0$ and no spin-glass phase for all $q$ (as is the case for the two-dimensional $\pm J$ EA model). 
Nevertheless, in two dimensions, the low temperature dynamics becomes extremely slow when decreasing the temperature and has some characteristics similar to a true spin-glass phase. 
This is the slow preasymptotic dynamical regime we will be interested in when analyzing the two-dimensional $\pm J$ Potts model in Secs.~\ref{sec:pmj} and \ref{sec:potts-glass}.

\subsection*{Quantities of interest and numerical implementation}

We use a Monte Carlo dynamical approach with Metropolis transition rates. The same simulation protocol 
is used for the Potts models with ferromagnetic and $\pm J$ couplings.
Starting from a completely disordered state (infinite temperature) we quench the system to a fixed working temperature $T$ at time $t=0$.
The key characterization tool we use is the MFTD. 
It measures the distribution of time scales associated to flipping events in the system, and it has proven to be a good quantity to expose time scale separation and thus dynamical heterogeneities. 
We measure the number of flips $N_F$ done by every spin between any two of the possible $q$ states in a time window $\Delta t = t - t_w$, where the waiting time $t_w$ corresponds to the time elapsed from the quench done at $t=0$. 
The mean flipping time formally depends on both $t$ and $t_w$ and is simply defined as $\tau = \Delta t/N_F$. 
From this quantity we construct the MFTD $P(\tau)$, which is typically shown as $P(\log_{10} \tau)$ due to its broadness. 
For clarity, the time evolution of the MFTD is followed using $\Delta t = t_w$. 
Results for other values of $\Delta t$ and $t_w$ are qualitatively similar.

In the last few years the use of graphics processing units (GPUs) to accelerate simulations has burst out in many areas of physics including fluid models, gravitation and superconductivity~\cite{Bernaschi2009,Herrmann2010,Covaci2011}.
In particular, in Statistical Physics many works have been devoted to report implementations of spin models on GPU architectures (see~\cite{Weigel2012,Weigel2011} and references therein).
These recent reports alert us about the tremendous benefit of working with GPU's implementations, which reduce significantly the simulation times.
For our numerical simulations in the present work, we use GPU-based parallel implementations of the Potts models. Among other benefits, the GPU-implementation of these models allowed us to work with system sizes up to $N=8192^2$ for the ferromagnetic model, and up to $N=16384^2$ for the $\pm J$ Potts model.
Averages where taken over $m$ thermal realizations for the ferromagnetic case and over $M$ disorder realizations for the $\pm J$ Potts case.
Details on the numerical implementation of the Potts model with ferromagnetic couplings can be found in Ref.~\cite{ferrero2012}. For further technical details see the Appendix~\ref{sec:Append}~\footnote{The present implementation of the $\pm J$ Potts model is available 
under GNU GPL 3.0 license at https://bitbucket.org/ezeferrero/potts-glass.}.

\section{Potts model with $q=9$}
\label{sec:coarsening}

\begin{figure}[!tbp]
\includegraphics[scale=0.35,clip=true]{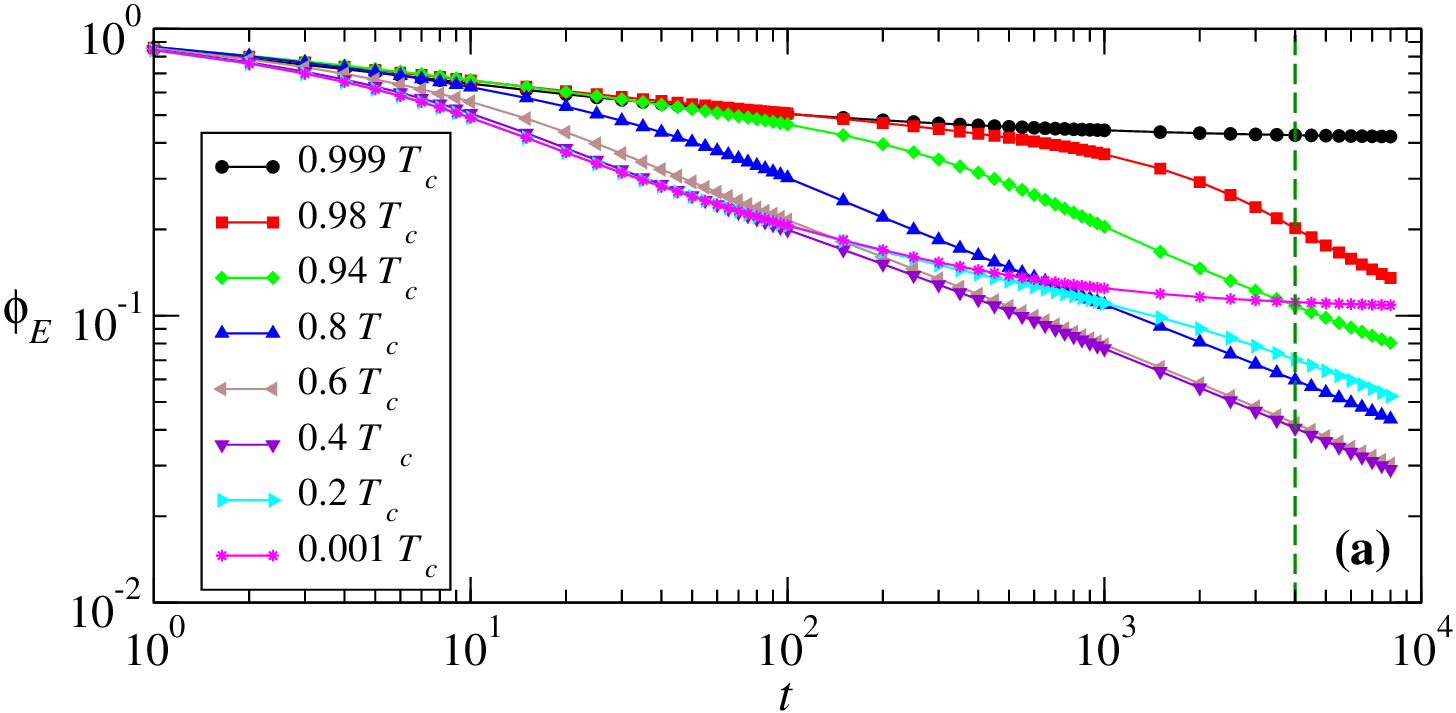}
\includegraphics[scale=0.35,clip=true]{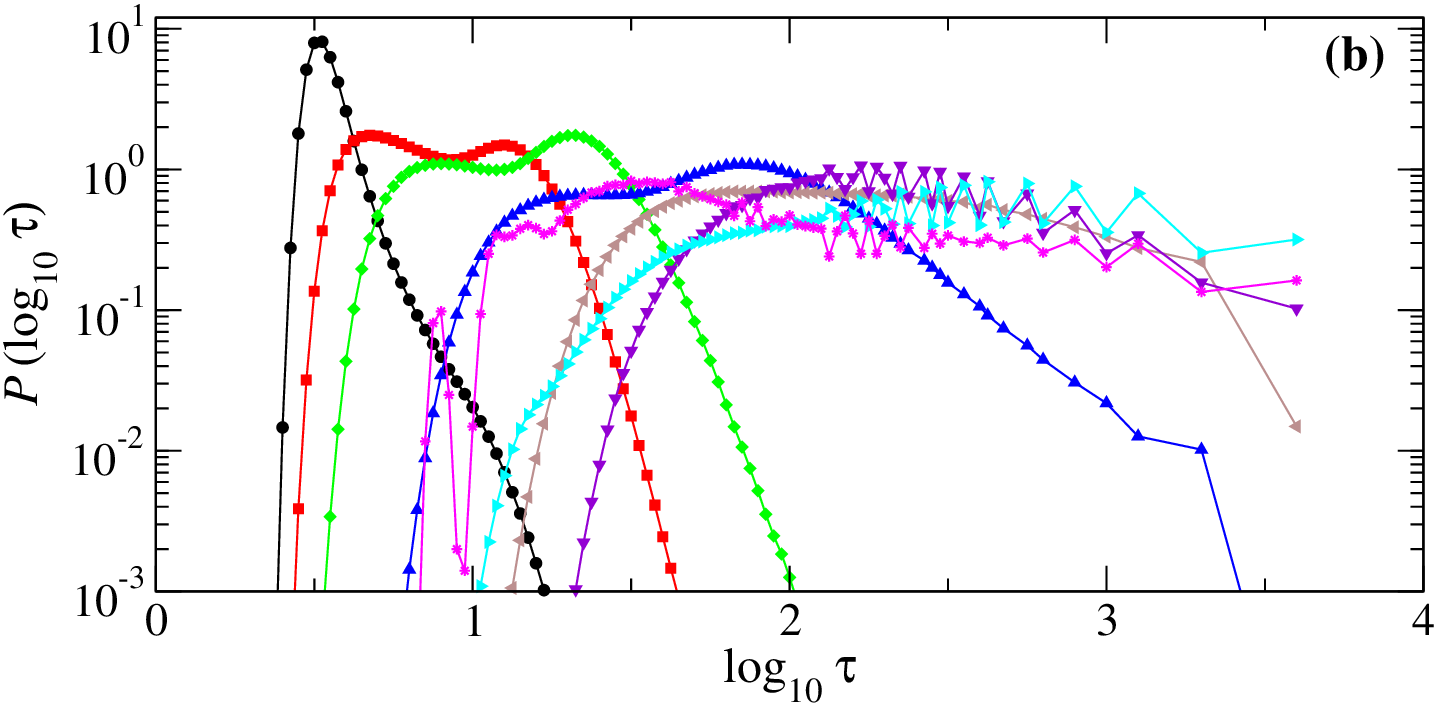}
\caption{\label{fig:Pottsq9-coarsening-temp} (Color online) 
(a) Relaxation of the excess energy for different temperatures for the $q=9$ Potts model.
(b) Dependence with temperature of the MFTD for the $q=9$ Potts model (colors and symbols for each temperature are conserved from the panel above).
All the curves correspond to $\Delta t = t_w = 4 \times 10^3$, $L=8192$, $m=100$.}
\end{figure}

In this Section we will analyze the MFTD for the $q=9$ Potts model for different dynamical regimes. 
When lowering the temperature below and close to $T_c$ the system rapidly passes the nucleation regime and enters the coarsening regime since $T_n \approx 0.995 T_c$.
This can be shown by the evolution of the relaxation function defined as the normalized excess of energy
\begin{equation}
 \phi_E(t)=\frac{e(t)-e(\infty)}{e(0)-e(\infty)},
\end{equation}
where $e(t)=\langle H \rangle/N$ is the average energy per spin and $e(\infty)$ is the equilibrium energy of the system.
Since this quantity relies on the value of $e(\infty)$ it is very difficult to obtain it for glassy systems, due to 
extremely long relaxation times. For this reason we have only analyzed $\phi_E$ for the characterization of the different 
dynamical regimes of the Potts model~\cite{ferrero2007}. 
Given a quench to a working temperature $T<T_c$ the relaxation function shows, after a short transient, an evolution depending on temperature. In a wide temperature range $0.2 T_c \lesssim T \lesssim 0.98 T_c$ one observes a power-law decay of the relaxation function $\phi_E \sim t^{-1/2}$, as shown in Fig.~\ref{fig:Pottsq9-coarsening-temp}(a).
This is consistent with a coarsening  dynamics where the size of the domains are growing as $\ell \sim t^{1/2}$ \footnote{Note that we are not reaching late times when $\ell \sim L$ and blocked states arise with a finite probability for $T<T_{co}$ and start to dominate the relaxation.}. 
Subsequently, the appearance of a glassy state slows down the dynamics, appreciated as a plateau in the relaxation function, which finally saturates in a finite value when $T \to 0$.

The MFTD is measured in the coarsening regime by choosing $\Delta t = t_w = 4 \times 10^3$. 
At temperatures close enough to $T_c$ the MFTD is expected to be characterized by a single peak at small time scales, typical of fast flipping events. 
Nevertheless, even at a temperature $T=0.999 T_c$ the MFTD shows a shoulder at larger time scales. 
From this shoulder, a well defined second peak develops when lowering the temperature inside the coarsening regime, as shown in Fig.~\ref{fig:Pottsq9-coarsening-temp}(b).
The interpretation of the first and second peak in the MFTD when the system is in the coarsening regime is clear: while the second peak is related to thermal excitations within one of the $q$--states ferromagnetic domains, the first peak is associated to the fast flipping events of spins belonging to domain walls between the different growing domains~\cite{roma2010}.

\begin{figure}[!tbp]
\includegraphics[scale=0.35,clip=true]{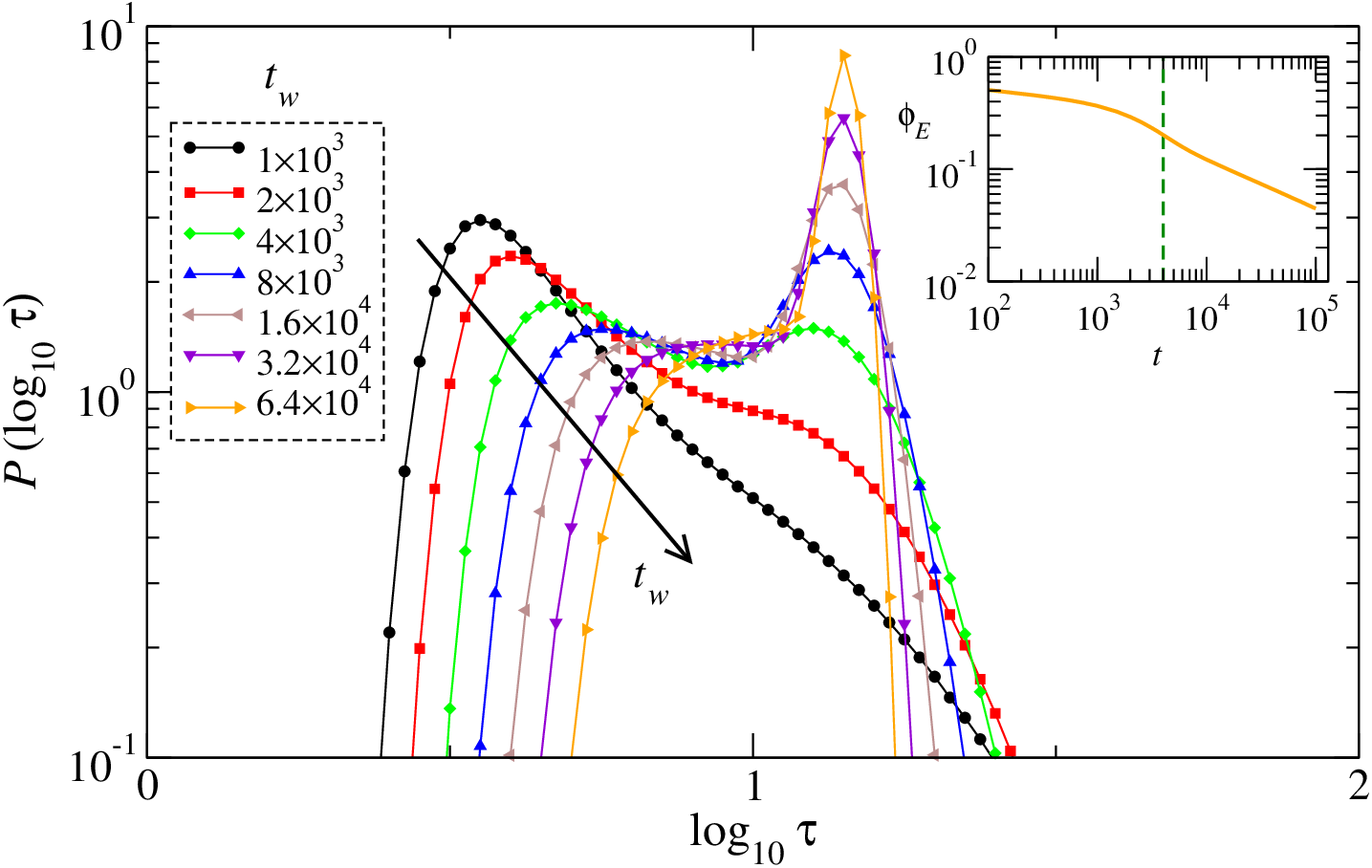}
\caption{\label{fig:Pottsq9-coarsening-time} (Color online)
Evolution with the waiting time of the MFTD for the $q=9$ Potts model at $T=0.98 T_c$ (within the coarsening regime). 
The time evolution is followed using a two-times protocol with $\Delta t=t_w$. 
Averages were taken with $L=8192$, $m=100$.
The inset shows the evolution of the relaxation function $\phi_E(t)$ at this temperature.
The vertical dashed line indicates approximately the beginning of the power law decay corresponding to the coarsening regime ($t \approx 4\times10^3$). 
}
\end{figure}

It is also possible to follow the time evolution of the MFTD at a given fixed temperature by changing $t_w$, using $\Delta t=t_w$. Figure~\ref{fig:Pottsq9-coarsening-time} shows the MFTD for different waiting times $t_w$ at $T=0.98 T_c$, which corresponds to the coarsening regime. As shown by the evolution of the relaxation function in the inset, the power-law decay begins at a time $t \approx 4 \times 10^3$. It is worth stressing that the evolution from a MFTD
with a single peak into a bimodal distribution starts manifesting before this time, nevertheless it is still clearly related to the coarsening process. 
The growth of the second peak with time indicates that the number of spins within ferromagnetic domains is increasing.
The first peak is not only decreasing but it is also moving to the right, indicating that flipping transitions are becoming harder for those spins participating in domain walls. It is also worth stressing that spins which belongs to a given ferromagnetic domain can eventually be part of a domain wall in a subsequent time, i.e. a given spin can contribute to either of the two peaks of the MFTD distribution while the coarsening process is developing.

When the temperature is lowered in the $q=9$ Potts model, the dynamics becomes slower as the different temperature scales are overpassed.
In this case a glassy state starts dominating the dynamics and the relaxation function is characterized by a \textit{plateau}~\cite{ferrero2007} (see Fig.~\ref{fig:Pottsq9-coarsening-temp}).
Due to the presence of this plateau, the second peak moves to larger flipping time values and eventually becomes difficult to characterize.
                                                                                                                                           
We have shown in this Section that the MFTD can be related to the domain growth process in the coarsening regime of the $q=9$ Potts model. 
In order to extend this analysis to a disordered model, and since it has been recently suggested that a coarsening process takes place within the backbone of the three-dimensional $\pm J$ EA model~\cite{roma2010}, in the following sections we will consider in our analysis the $\pm J$ Potts model. 
As a starting point we will consider the $\pm J$ $q=2$ case, i.e. the $\pm J$ EA model. Then, we will also consider larger values of $q$ which shows the importance of considering of the backbone structure in Potts models.

\section{$\pm J$ EA model}
\label{sec:pmj}

In this section we move to the study of disordered systems first analyzing the $\pm J$ EA model. We first briefly discuss the results already found for the $\pm J$ EA model in relation with dynamical heterogeneities and the backbone structure, which is defined based on information given by the topology of the ground state~\cite{roma2010b}. Concretely, those bonds which do not change their state -- satisfied or frustrated -- in all configurations of the multiply degenerated ground state of the model compose the rigid lattice. Besides, those spins connected through the rigid lattice are labeled as solidary spins, since they do maintain their relative orientation in all ground state configurations. The rest of the spins are labeled as non-solidary spins. Both the rigid lattice and the set of solidary spins form the backbone structure of the system.

In Ref.~\cite{roma2010} the three-dimensional $\pm J$ EA model was studied and slow and fast time scales were related to the spatial heterogeneities given by the backbone structure. 
Evidence was presented for a growing process taking place inside the backbone structure, where spins can be ordered in a ferromagnetic-like state.
In this case, the backbone structure has a finite component percolating all over the sample where ferromagnetic-like correlations can grow~\cite{roma2010b}.
The situation is different in the two-dimensional $\pm J$ EA model~\cite{Roma2,roma2010b}, where the backbone structure is fragmented and does not percolate through the sample.
However, it has been shown that the distribution of islands of the backbone structure is very close to the one corresponding to the percolation threshold~\cite{roma2010b}.
If one considers the possibility of growing ferromagnetic-like correlations inside each fragment of the backbone structure in the two-dimensional case, then a growing process would take place with a cut off given by the non-percolative character of the backbone structure.
In the following we further test this idea.

\begin{figure}[!tbp]
\includegraphics[scale=0.35,clip=true]{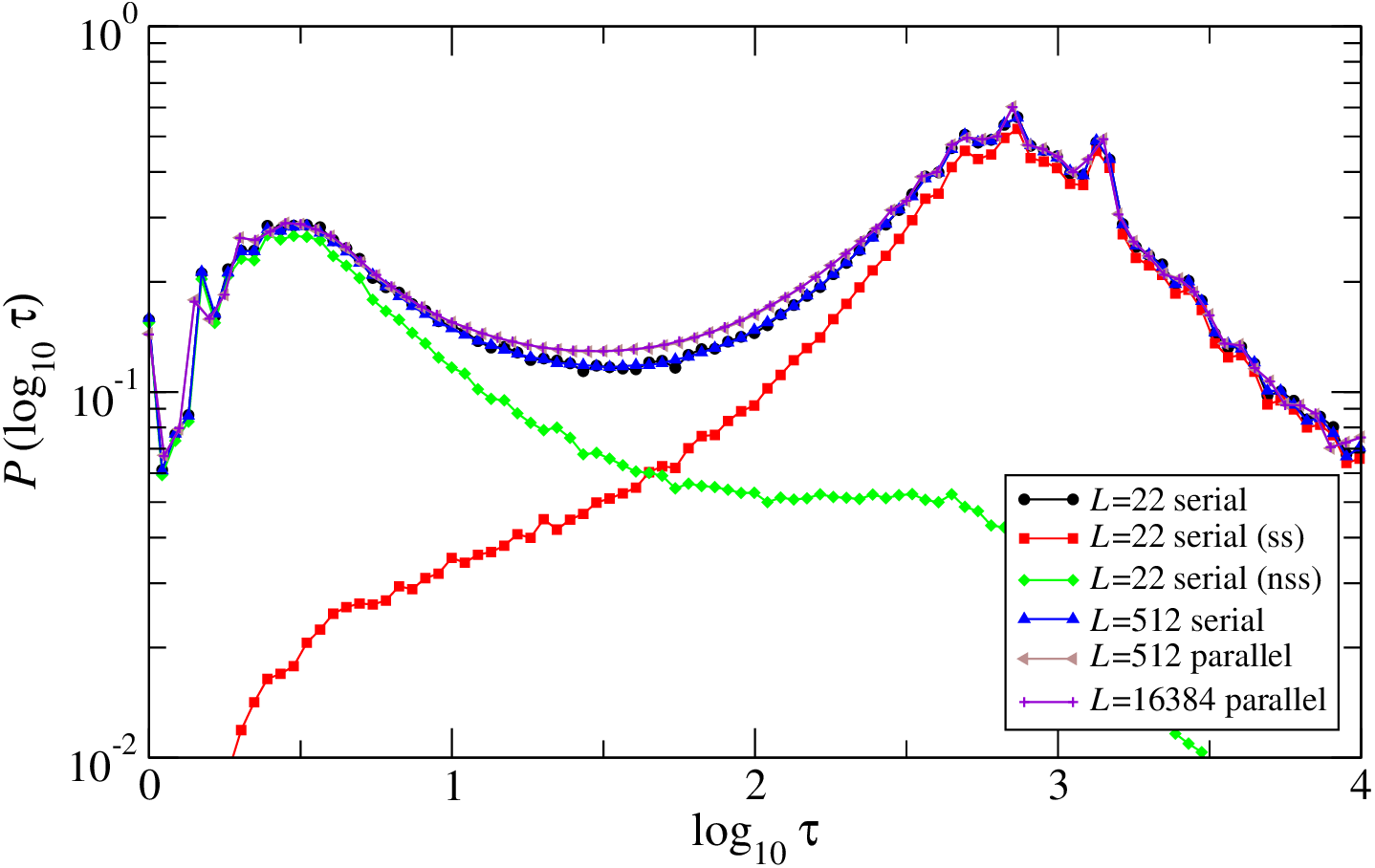}
\caption{\label{fig:EA2Dq2-dynamics} (Color online) 
MFTD for the $\pm J$ EA model (equivalently $\pm J$ Potts model with $q=2$) in two dimensions for $T=0.25 J$ and $\Delta t=t_w=10^6$. 
We show the full MFTD for $L=22$ and $L=512$ using spin flip dynamics with serial random updates together with the result 
for $L=512$ using the parallel dynamics described in Sec.~\ref{sec:model}. 
The separation of the full MFTD into the contributions from solidary spins (ss) and non-solidary spins (nss) is also shown for $L=22$.
Averages were taken over $M=2000$ samples for $L=22$ and $M=100$ samples for $L=512$.
}
\end{figure}

\begin{figure}[!tbp]
\includegraphics[scale=0.35,clip=true]{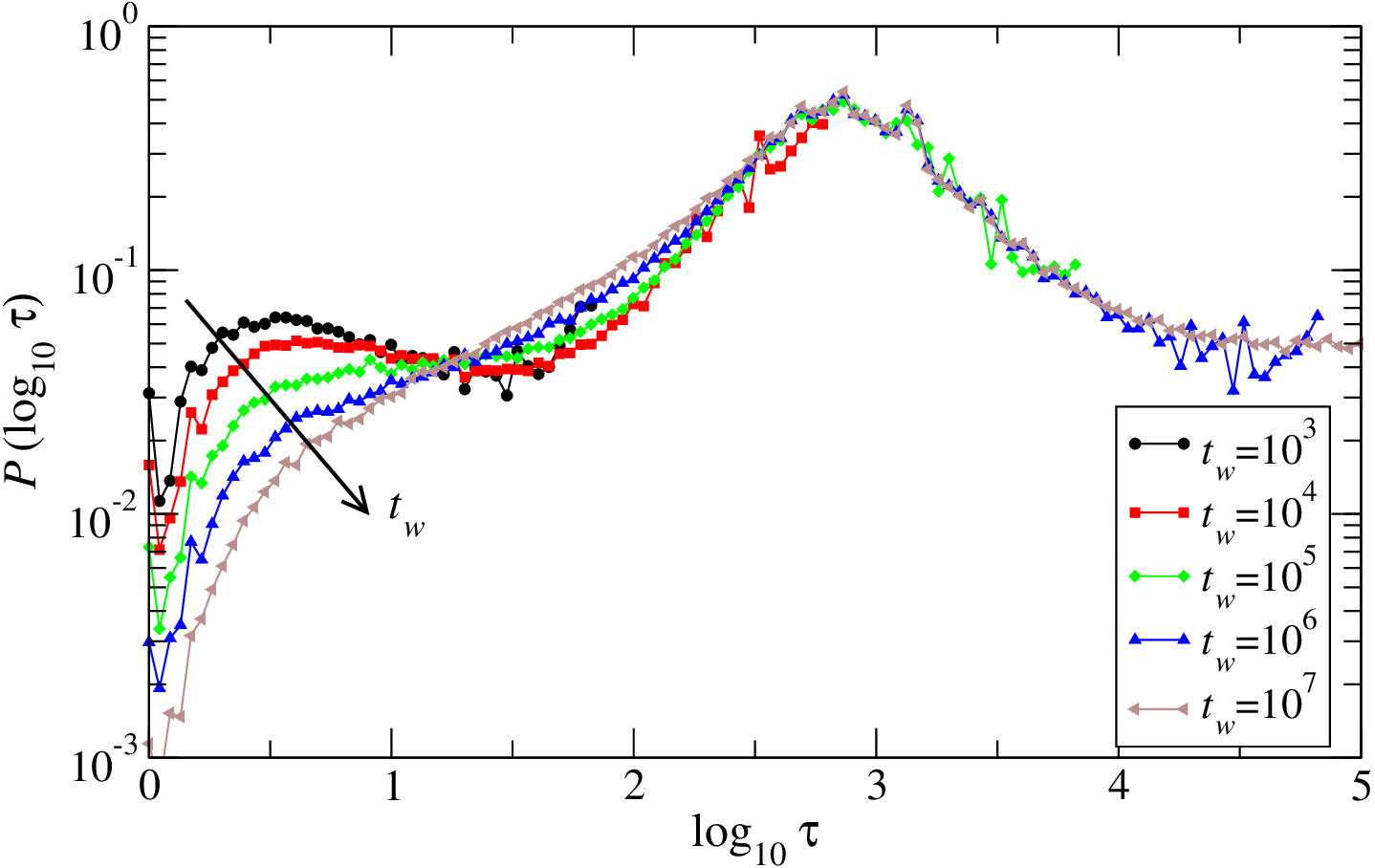}
\caption{\label{fig:EA2Dq2-solidary} (Color online) 
Evolution of the MFTD given by the solidary spins in the two-dimensional $\pm J$ EA model. 
Each curve was obtained using $\Delta t=t_w$. The temperature is $T=0.25J$ in all cases. 
The arrow emphasizes how the contribution to the fast peak is evolving with increasing $t_w$.
Averages were taken within $M=2000$ samples for $L=22$.
}
\end{figure}

First, in Fig.~\ref{fig:EA2Dq2-dynamics} we show the full MFTD and its separation into the contribution of solidary and non-solidary spins, according to the backbone structure obtained from ground state information.
In this case $\Delta t = t_w =10^6$ and the temperature is $T=0.25J>T_c=0$, at which the dynamics is sufficiently slow, i.e. in the preasymptotic glassy regime.
The separation is reported for the linear size of the system $L=22$, larger than the one reported in Ref.~\cite{Roma2}.
Again this separation perfectly accounts for the observed dynamical heterogeneities.

The numerical simulations showing the separation of the MFTD into its contributions from solidary and non-solidary spins for $L=22$ were performed using single spin flip dynamics with Metropolis transition probabilities and random updates.
Moreover, the full MFTD is compared in Fig.~\ref{fig:EA2Dq2-dynamics} with the one for $L=512$ obtained with the same dynamical rules, showing a considerable agreement. This indicates that although $L=22$ seems to be a rather small system size it already perfectly accounts for the main dynamical properties of the system. One can also compare this result with the one obtained using 
the GPU-based parallel implementation. The MFTD for $L=512$ and $L=16384$ (almost indistinguishible between them)  obtained with this dynamical rules are also presented in Fig.~\ref{fig:EA2Dq2-dynamics} and the agreement with the serial implementation is excellent. This validates the use of this dynamical rules for the present dynamical studies, i.e. there are not significant differences between serial and parallel spin flip updates.

Finally, we show in this section the evolution of the contribution of solidary spins to the MFTD corresponding to the two-dimensional $\pm J$ EA model.
We show results for $T=0.25 J$ and in the range $10^3 < \Delta t = t_w < 10^7$, where the full MFTD remains mostly unchanged.
From the evolution of the contribution of solidary spins to the MFTD presented in Fig.~\ref{fig:EA2Dq2-solidary}, it can be observed that while the second peak remains  almost unchanged, the first peak is disappearing (or moving to the right).
This can be taken as an indication of the presence of a growing process inside the backbone structure as previously reported for the three-dimensional case~\cite{roma2010}. Note also the similarities between this result and the one obtained for the Potts model with ferromagnetic couplings in Fig.~\ref{fig:Pottsq9-coarsening-time}. The second peak of the contribution of solidary spins to the MFTD, corresponding to slow degrees of freedom, is not evolving with $t_w$. Indeed, the second peak only depends on temperature an is thermally activated~\cite{Roma2}, as further discussed below.

\section{$\pm J$ Potts model}
\label{sec:potts-glass}

\begin{figure}[!tbp]
\includegraphics[scale=0.35,clip=true]{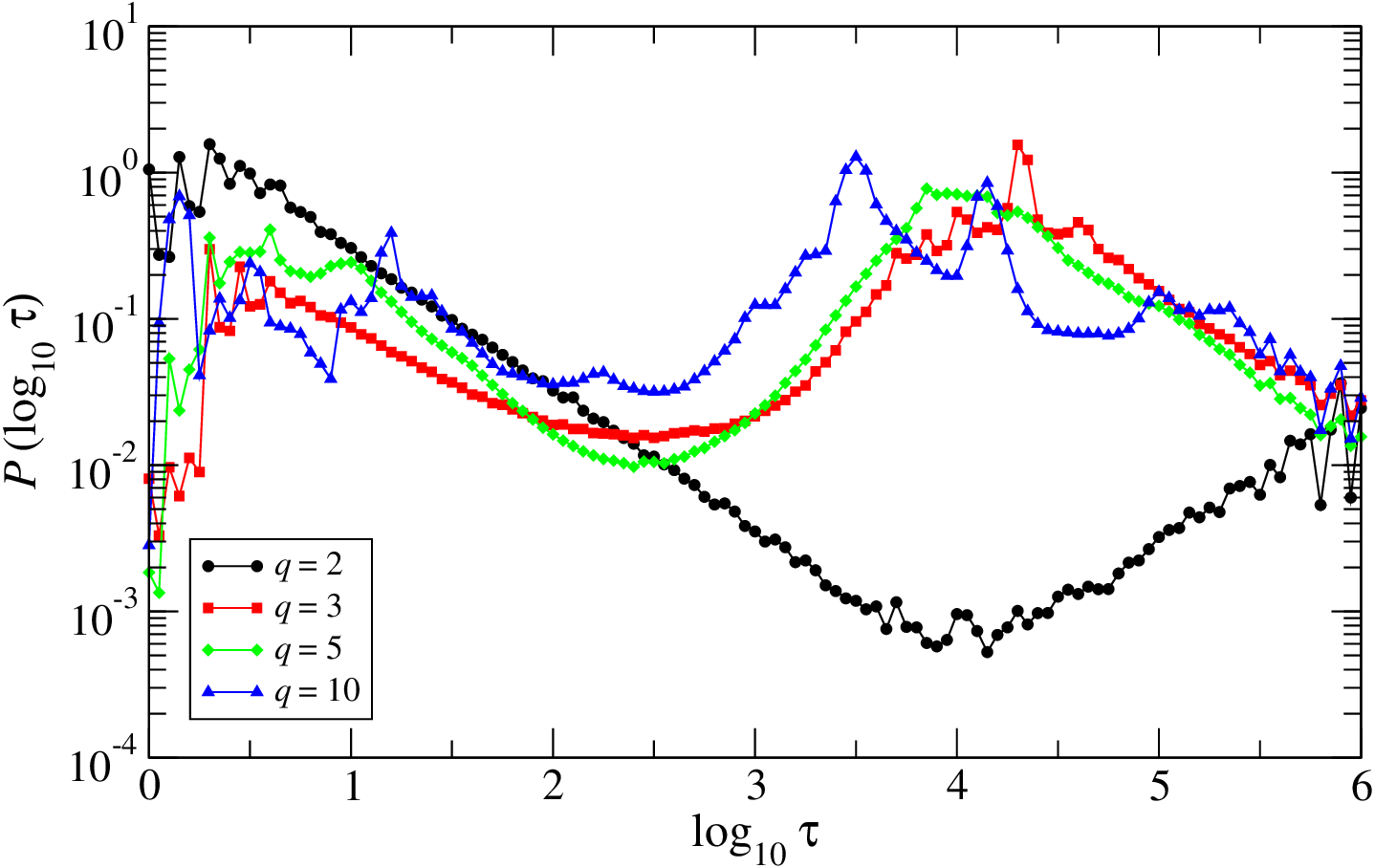}
\caption{\label{fig:PottspmJ-qvalues} (Color online)
MFTD for the $\pm J$ Potts model with different $q$--states, $q=2,3,5,10$, at a fixed temperature $T=0.1J$. 
Curves correspond to $\Delta t=t_w=10^7$, $L=4096$ and $M=10$.
}
\end{figure}

In this section we extend the study of the MFTD to the two-dimensional $\pm J$ Potts model with different $q$ values using GPU-based numerical simulations.
We will compare the full MFTD for different $q$ values and discuss its temperature dependence.
Figure~\ref{fig:PottspmJ-qvalues} shows the MFTD for the $\pm J$ Potts model with different number of available states $q=2,3,5,10$ for each spin and using $T=0.1J$ and $\Delta t = t_w = 10^7$. 
One can observe that at this low temperature the MFTD can be decomposed on at least two peaks, thus revealing the strong nature of the observed heterogeneous dynamics. 
Comparing the curve for $q=2$ showed in this figure ($T=0.1J$) with the one showed in Fig.~\ref{fig:EA2Dq2-dynamics} ($T=0.25J$), we see that the second peak is moved out to the right due to thermal activation. However, for $q=3,5,9$ the time scale of the second peak is within the time window.
For values of $q>2$ one can envisage the presence of more than two peaks, as is more clear for $q=10$, suggesting a more complex underlying structure. 

\begin{figure}[!tbp]
\includegraphics[scale=0.35,clip=true]{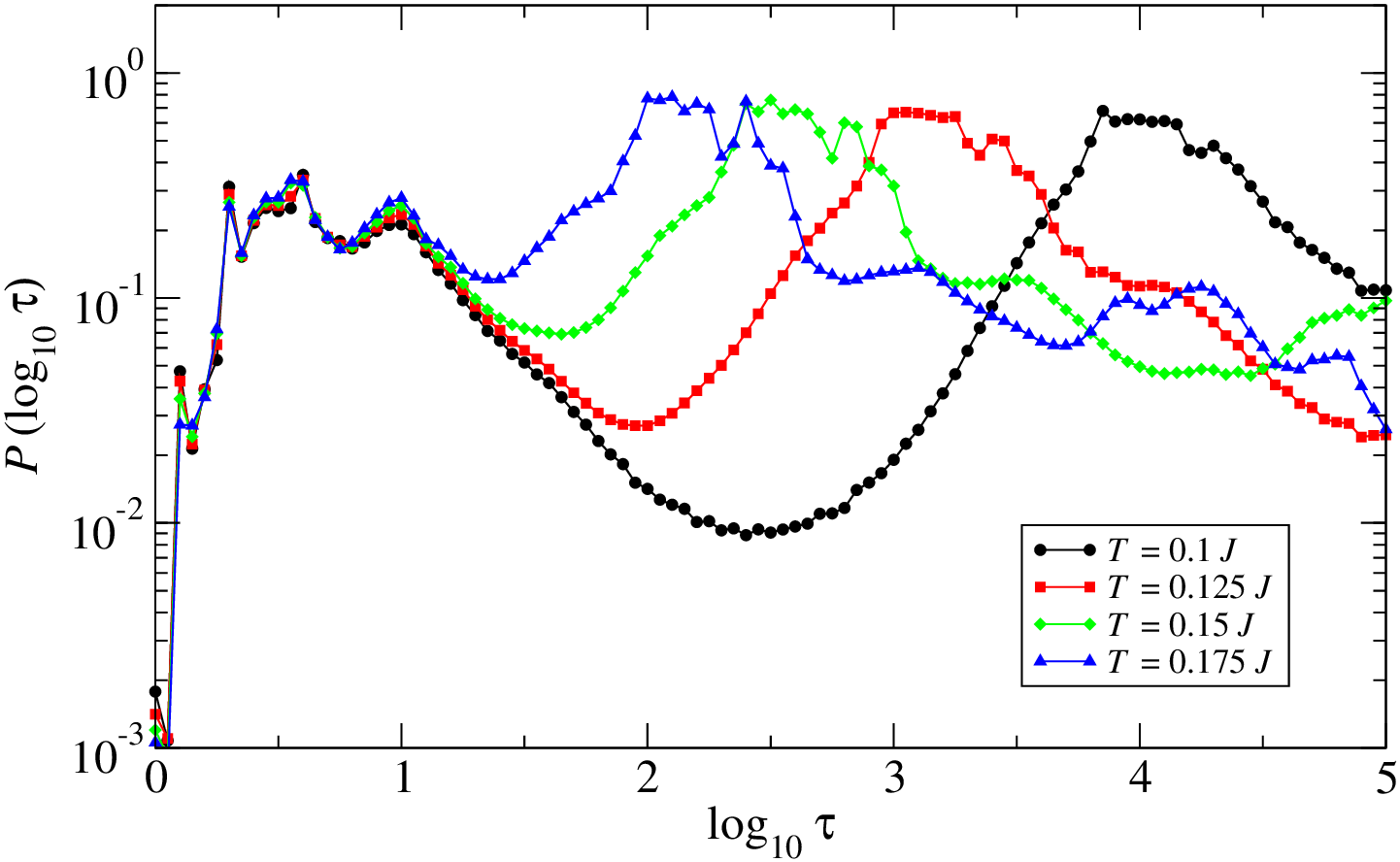}
\caption{\label{fig:PottspmJ-peak-q5} (Color online) 
MFTD for the $\pm J$ Potts model model with $q=5$, $\Delta t=t_w=10^7$, and different temperatures as indicated.
Curves correspond to $L=2048$, $M=10$.}
\end{figure}

\begin{figure}[!tbp]
\includegraphics[scale=0.35,clip=true]{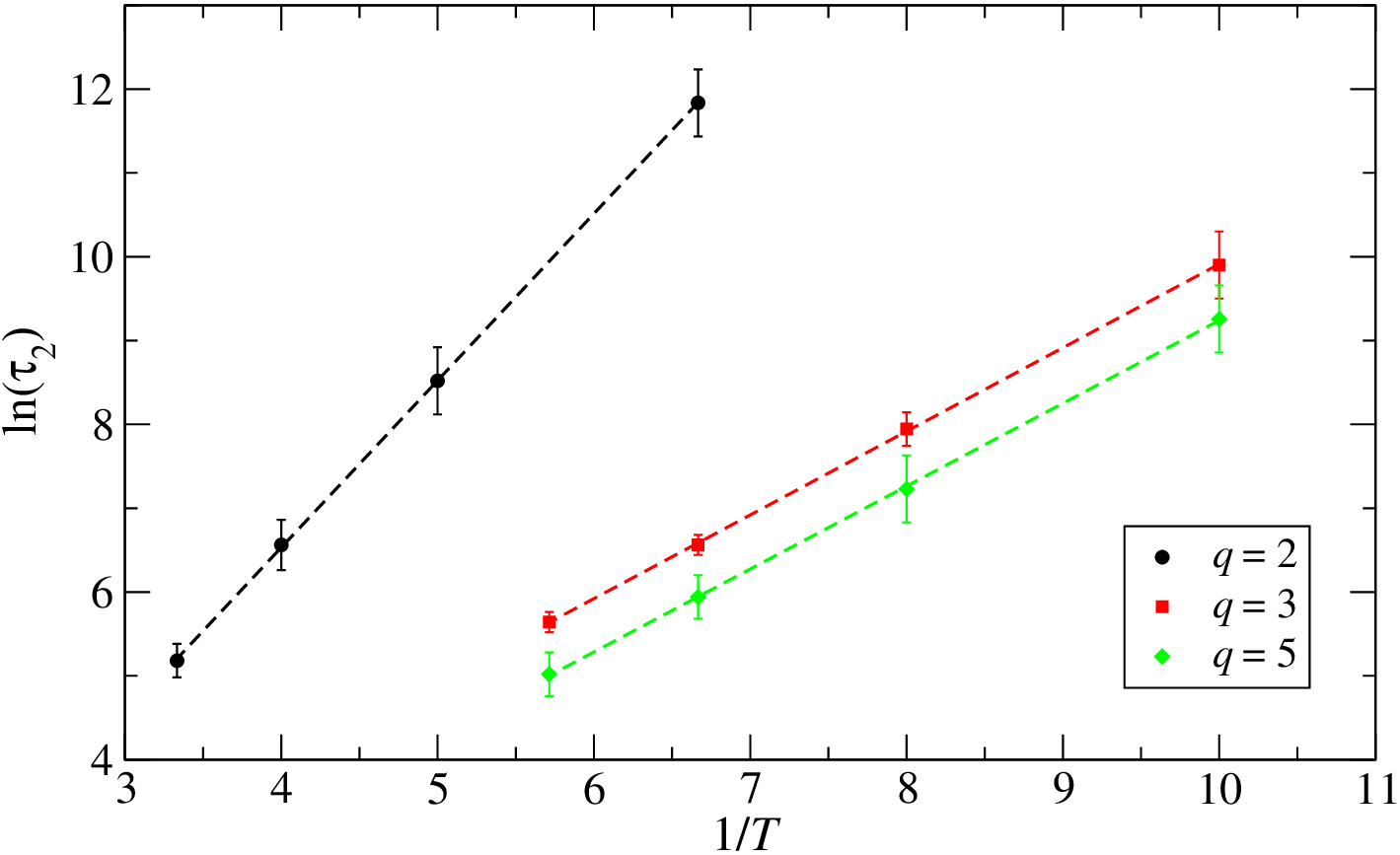}
\caption{\label{fig:PottspmJ-peak-activation} (Color online) 
Arrhenius plot for the characteristic time scale of the second peak ($\tau_s$) of the MFTD for $q=2,3,5$ ($\Delta t=t_w=10^7$).
The dashed lines are linear fits of the data (symbols), giving slopes of $1.99\pm0.01$ for the $q=2$ case and $0.99\pm0.01$ for $q=3$ and $q=5$ cases.}
\end{figure}

The existence of the strong time-scale separation observed in Fig.~\ref{fig:PottspmJ-qvalues} can be attributed to the presence of an underlying backbone structure.
In order to further pursue this idea we show in Fig.~\ref{fig:PottspmJ-peak-q5} the temperature dependence of the MFTD for $q=5$ and $\Delta t = t_w = 10^7$.
The first peak characterizing fast degrees of freedom does not change appreciably with temperature.
The second peak, related to the slow degrees of freedom, moves to larger values of the mean flipping time scale when the temperature is decreased.
Indeed, the average position of the second peak can be thought of as a characterization of the states with lowest excitation energies in the system.
Figure~\ref{fig:PottspmJ-peak-activation} shows that the average position of the second peak, $\tau_2$, is thermally activated for different $q$ values ($q=2,3,5$).
The activation energy is $\Delta E = 2J$ for $q=2$ and $\Delta E = J$ for $q>2$.
This can be understood by inspection of the Hamiltonian~\eqref{eq:H-Potts-glass} presented in Sec.~\ref{sec:model}.
While for the $q=2$ case the lowest excitations correspond to flipping one spin with only one frustrated bond, for $q>2$ the lowest excitation is given by flipping one spin with two frustrated bonds. In both cases the final excited state has three frustrated bonds.
This therefore clearly shows that the structure around the second peak is thermally activated, which we take as another indication of a growing process inside the backbone structure.

The dynamical properties described here for the $\pm J$ Potts model implicitly point to the presence of a set of strongly correlated spins, reminiscent of the set of solidary spins in the $\pm J$ EA model. The presented evidence therefore suggest the existence of a backbone structure in the $\pm J$ Potts model. This backbone structure would be composed of solidary spins of a ferromagnetic-like character and non-solidary spins with paramagnetic-like features.
Interestingly, although we have presented evidence for the presence of a backbone structure in the $\pm J$ Potts model, it is not straightforward to compute it from ground state information. In fact, due to the internal variable $q$ in each spin, the protocol to obtain the backbone structure used for $q=2$ should be generalized. This task, which is clearly far from trivial, will be left for future work.

\section{Discussions and concluding remarks}
\label{sec:conc}

In the present work we have studied the dynamical heterogeneities characterized through the MFTD in the Potts model with both ferromagnetic and $\pm J$ couplings.
We have also devoted our analysis to relate these dynamical heterogeneities with the presence of a backbone structure.

First, we showed how the MFTD accounts for the different dynamical regimes observed in the $q=9$ Potts model.
In particular, by analyzing the system in the coarsening regime one can give support to the idea that the temperature and time evolution of the two peaks 
in the MFTD are intimately related to the coarsening of different domains.

We showed that for the two-dimensional $\pm J$ EA model (equivalently the $\pm J$ Potts model with $q=2$), where the backbone structure is well characterized, the time evolution of the MFTD strongly suggests that a growth process is taking place within the preasymptotic regime.
We have also analyzed the MFTD for the $\pm J$ Potts model with $q>2$.
We have shown numerical evidence suggesting that this model also has an underlying backbone structure.
In particular the thermal activation of the time scale related to the second peak indicates that a backbone structure can be the key ingredient dominating the glassy dynamics. As a by-product, we have also shown the equivalence between serial and parallel implementations of the numerical simulations, a fact that can not be assumed \textit{a priori}.

Since we have presented numerical evidence for the plausibility of the existence of a backbone structure in the $\pm J$ Potts model, the question of how to identify this backbone structure naturally arises.
The most trivial test one can think of is to use the backbone structure obtained for $q=2$ to analyze the dynamical properties for $q>2$.
In this way one can test whether the spin degrees of freedom are relevant or not in defining the backbone.
We have run such test and we observed that the rigid lattice defined in the two-dimensional $\pm J$ EA model does not give a proper time scale separation for $q>2$.
From this simple test we can assure that the backbone structure is a non-trivial combination of the spin degrees of freedom and the couplings distribution.

\begin{acknowledgments}
The authors thank fruitful discussions with N. Wolovick and C. Bederi\'an.
We would like to thank to the GTMC group of FaMAF-UNC for kindly giving access to their cluster with GPUs, for benchmarks and simulations.
We also acknowledge an NVIDIA Academic Partnership Program granted to D. Dom\'{\i}nguez at Centro At\'omico Bariloche.
FR acknowledges financial support from  projects PIP114-201001-00172
and PICT07-2185, and U.N. de San Luis under project 322000.
SB is partially supported by projects PICT2010-889 and PIP11220090100051.
\end{acknowledgments}

\appendix*
\section{GPU-based numerical implementation}
\label{sec:Append}

For our numerical simulations we use GPU-based parallel implementations of the Potts models. In particular a checkerboard scheme with parallel 
spin flip updates is used. It is worth stressing that although equilibrium measures are accepted as independent of the particular Monte Carlo implementation, this is not usually the case for the dynamical properties of a model~\cite{Landau2000}. In principle, a parallel implementation of 
a stochastic dynamics could introduce undesirable correlations, and thus should be tested beforehand. 
Here, we have checked that our GPU-based massively parallel implementation is completely compatible with the usual serial CPU-based dynamics (see Sec.~\ref{sec:pmj}) where spins are chosen at random to attempt each update, and this was done for a totally out-of-equilibrium quantity such as the MFTD.

On the base of the code presented in Ref.~\cite{ferrero2012} some modifications were implemented to allow for the construction of the MFTD histograms and introduce competitive interactions in the $\pm J$ Potts model. For the non-disordered case we have just added a few lines to the update routine to keep count of the number of flips of each spin in the time window we are interested in, and a routine to calculate the $P(\tau)$ histograms 
using that information. Since an additional write operation to an array in global memory is necessary at each local update step, we get a slowdown of 5 to 10\% with respect to the reference code, at an average spin-flip-time of $0.2$ns. Is it worth stressing that we are still simulating at more than 120x with respect to a CPU implementation.

For the $\pm J$ Potts model the extension is straightforward.
In a naive implementation, we have stored the $J_{ij}$ bonds in an array in global memory.
Every spin has direct access to the value of the four bonds shared with its nearest neighbors. Even more, the information of the bonds is duplicated in global memory.
This array is built in such a way that neighbors threads can copy down the values of their bonds to register accessing in a coalesced manner.
This implementation suffers a slowdown of 16\% with respect to the non-disorder case, but still has an acceptable performance (0.22ns per spin flip on a GTX 480 \footnote{Compiler nvcc 4.1 V0.2.1221 options \emph{-O3 -arch=sm\_20 --use\_fast\_math}}) for a spin-glass code as compared with other implementations of realistic system sizes~\cite{Yavorskii2012}. As compared with the C code used for the serial numerical simulations of the $\pm J$ EA model, the present GPU-based implementation represents a speed up of 60x.
In both cases, to calculate the MFTD a routine to build the histograms is implemented using atomic operations on shared memory.

The implementation of a random number generator in GPU should be carefully chosen (for a recent review see~\cite{Manssen2012})  
As in Ref.~\cite{ferrero2012} we have used the Multiply-With-Carry random number generator which allows for the simultaneous generation and use of several independent random sequences, and has a good trade-out of performance and statistical quality, at least for spin systems implementations.

\bibliography{spinglass,structural-glass}
\bibliographystyle{apsrev4-1}

\end{document}